%% file: ipin_2018_cz_aw.tex
\documentclass[conference]{IEEEtran}
\usepackage{amsmath}
\usepackage{amsfonts}
\usepackage{amssymb}
\usepackage{mathptmx}       % selects Times Roman as basic font
\usepackage{makeidx}         % allows index generation
\usepackage{graphicx}        % standard LaTeX graphics tool
\usepackage{multicol}        % used for the two-column index
\usepackage{acronym}
\usepackage{multirow}
\usepackage{array}
\usepackage{color, soul}

\usepackage[caption=false, font=footnotesize]{subfig}
\usepackage{setspace}

\usepackage{footnote}
\makesavenoteenv{tabular}
\makesavenoteenv{table}
\usepackage{nameref}
\usepackage{cite}
\usepackage{fancyhdr}
\fancypagestyle{firstpage}{%
	\chead{2018 International Conference on Indoor Positioning and Indoor Navigation (IPIN) 2018, Nantes, France}
}

\makeindex             % used for the subject index
\DeclareMathAlphabet{\mathcal}{OMS}{cmsy}{m}{n}
\SetMathAlphabet{\mathcal}{bold}{OMS}{cmsy}{b}{n}

\begin{document}
	\pagenumbering{gobble}
	%input the defined acronyms
%	\input{../miscellaneous/cz_acronyms.tex}
%	\input{../miscellaneous/cz_command.tex}
	\input{cz_acronyms.tex}
	\input{cz_command.tex}
	%	\makesavenoteenv{tabular}
	%	\makesavenoteenv{table}
	
	\title{CDM: Compound dissimilarity measure and an application to fingerprinting-based positioning}
	%\hl{|}
	%	\author{Caifa~Zhou,
	%		Andreas~Wieser~\IEEEmembership{Member,~IEEE}% <-this % stops a space
	%		\thanks{This paper is extended from \cite{Zhou2018}, a paper published in LBS 2018, Zurich.}
	%		\thanks{C. Zhou and A. Wieser are with the Institute of Geodesy and Photogrammetry, ETH Zurich, 8093 Zurich, Switzerland, e-mail: caifa.zhou@geod.baug.ethz.ch; andreas.wieser@geod.baug.ethz.ch.}% <-this % stops a space
	%		\thanks{Manuscript received ....; revised .....}}
	\author{
		\IEEEauthorblockN{Caifa Zhou,  Andreas Wieser}
		\IEEEauthorblockA{IGP, ETH Z\"urich\\
			Stefano-Franscini-Platz 5, 8093 Z\"urich, Switzerland\\
			\{caifa.zhou, andreas.wieser\}@geod.baug.ethz.ch}
	}
	\maketitle
	
	\thispagestyle{firstpage}
	%\doublespacing
	\begin{abstract}
		A non-vector-based dissimilarity measure is proposed by combining vector-based distance metrics and set operations. This proposed \acf{cdm} is applicable to quantify the similarity of collections of attribute/feature pairs where not all attributes are present in all collections. This is a typical challenge in the context of  \eg \acf{fbp}. Compared to vector-based distance metrics (\eg Minkowski), the merits of the proposed \acs{cdm} are i) the data do not need to be converted to vectors of equal dimension, ii) shared and unshared attributes can be weighted differently within the assessment, and iii) additional degrees of freedom within the measure allow to adapt its properties to application needs in a data-driven way. 
		
		We indicate the validity of the proposed \acs{cdm} by demonstrating the improvements of the positioning performance of fingerprinting-based \acs{wlan} indoor positioning using four different datasets, three of them publicly available. When processing these datasets using \acs{cdm} instead of conventional distance metrics the accuracy of identifying buildings and floors improves by about 5\% on average. The 2d positioning errors in terms of \acf{rmse} are reduced by a factor of two, and the percentage of position solutions with less than 2$ \text{m} $ error improves by over 10\%.
	\end{abstract}
	
	% Note that keywords are not normally used for peer review papers.
	
	\section{Introduction}
	\label{sec:introduction}
	The core idea of this paper comes from analyzing a particular challenge occurring during the online step of a fingerprinting-based indoor positioning system (\eg using the \acf{rss} from \acs{wlan} \acfp{ap} as the features) based on the nearness in the fingerprint space as the principle for localization (\eg \acs{knn}). Each measured fingerprint consists of a collection of actually observed attributes (\eg identifications of \acsp{ap} and corresponding signal strengths). Fingerprints measured at different locations or at different times may contain different numbers of measurable \acsp{ap} \eg due to changed availability of \acsp{ap} or changed signal reception conditions. In such cases it is not clear how the similarity/dissimilarity between such fingerprints should be assessed, and in particular, how the similarity/dissimilarity between a fingerprint measured online (i.e. when the user position is to be determined) and the fingerprints collected in the \acf{rfm}, a collection of fingerprints with labeled ground locations for representing the functional relationship between the location and fingerprint, should be handled.
	
	In a general context this belongs to the class of missing data problems \cite{doi:10.1080/01621459.1994.10476768} mostly addressed in the fields of data analysis, data mining and machine learning \cite{doi:10.1002/9781118521373.wbeaa147}. A comprehensive review of the missing data problem is out-of-scope of this paper. Instead we focus on a concrete proposal to handle this problem within the context of positioning. In previous publications the authors either formulated the online measurements into vectors of equal length \cite{Padmanabhan2000, he2016wi,7565565} or used only the measurability of the individual attributes as binary features \cite{6071929}. The former scheme requires filling in values for missing attributes and ignoring newly measured ones. In this way it is easy to apply vector-based distance metrics for computing the dissimilarity but there are two disadvantages. One is the limited flexibility in dealing with missing or newly available data, the other one is time and computational resource cost: in most cases the number of all \acsp{ap} contained in the \acs{rfm} is much larger than the number of \acsp{ap} in an individual measured fingerprint. Therefore the vectorized data of uniform dimension which need to be composed and handled typically have many more elements than the individual measurements. The approach mapping the measured \acsp{ap} into a set of binary features, instead, is efficient in the sense of computational burden for assessing dissimilarity but it does not take the actual similarity of the measured values into account and thus does not support exploiting the potential for accurate positioning.
	
	After analyzing how this case is handled in previous publications, we explore the possibility of estimating the dissimilarity between the measurements which have the characteristics of partially missing observations of the attributes without formulating them into vectors of equal length. To this end we propose a non-vector-based dissimilarity measure (\secPref\ref{sec:proposedApproach}) which is a compound of a typical distance metric (\eg Minkowski) and set operations. In addition, we exploit the applicability of the proposed \acf{cdm} by applying it to four datasets used for \acf{fbp} and the result proves the benefits of the proposed dissimilarity measure (\secPref\ref{sec:experimentalResults}).
	
	\section{Related work of distance metrics}\label{sec:related}
	
	The concept of distance metrics used for measuring the nearness between the online measured features and the ones stored in the \acs{rfm} is a key for the realization of \acs{fbp} algorithms. The Euclidean distance is one of the most prevalent metrics in different research fields and communities \cite{cha2007comprehensive}. However, there is a variety of alternative distance metrics which may be more suitable for certain applications. In \cite{cha2007comprehensive}, Cha reported over 40 distance metrics or measures and analyzed their capability of measuring the difference between \acfp{pdf}. Minaev et. al. \cite{8115922} followed Cha's research  and applied them to an \acs{fbp} algorithm \acs{knn} by using the synthetic \acs{rss} from \acs{wlan} \acsp{ap} as the fingerprints and found that Lorentzian distance performs best among them. In \cite{TORRESSOSPEDRA20159263}, Torres-Sospedra et. al. surveyed and analyzed the performance of different distance metrics by applying them to a fingerprinting-based \acs{wlan} indoor positioning system which covers multi-buildings and multi-floors (\ie \textit{UJIIndoorLoc} dataset). In this paper, we propose the concept of \acs{cdm} joining it with the 8  distance metrics (see \tabPref\ref{tab:metrics}) performing best according to \cite{8115922}, and apply \acs{cdm} to  \acs{fbp} using four different datasets.
	
	\section{Compound dissimilarity measure}\label{sec:proposedApproach}
	% Always give a unique label
	% and use \ref{<label>} for cross-references
	% and \cite{<label>} for bibliographic references
	% use \sectionmark{}
	% to alter or adjust the section heading in the running head
	
	Suppose that there are several collections of measurements which need to be compared but differ with respect to the number and type of data included. For instance, the measurements in a WLAN-based indoor positioning system consist of the \acsp{rss} from all available \acsp{ap} at individual locations. However, only \acsp{ap} associated with \acs{rss} values exceeding the measuring sensitivity of the used WiFi device are observable at the individual locations, thus the \acsp{ap} measured at different points in space time will differ. Each measurement consists of an attribute \eg the \acf{mac} address of the respective \acs{ap}, and an \acs{rss} value. The fingerprint at a particular location is the collection of measurements actually made at that location. As stated above, we propose an approach herein to estimate the similarity or dissimilarity between such fingerprints without reformulating all the measurements with different attributes into vectors of equal length as in several other publications \cite{he2016wi, mautz2011survey,Zhou2018,Padmanabhan2000}.
	
\begin{figure}[!htb]
		\centering
		\includegraphics[width=0.6\columnwidth]{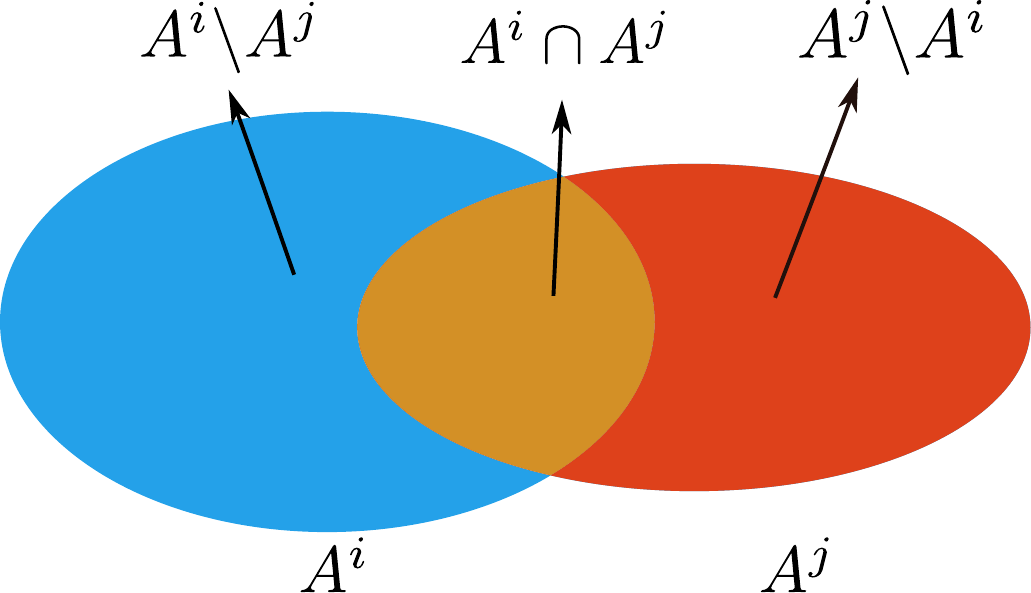}
		\caption{The scheme of calculating the dissimilarity measure into three parts.}
		\label{fig:scheme_three_parts}
	\end{figure}

	Given $ n $ measurements denoted as $ \{\mathbf{O}^i\}_{i=1}^{n} $, each measurement consisting of a set of paired attribute and measured value (\eg the \acs{rss}), \ie $ \mathbf{O}^i:=\{(a, {v}^i_a)| a \in A^i,  {v}^i_a\in \mathcal{R}\}, i=1,\cdots,n $, where $ {A}^i \subseteq \mathcal{M}$ is the set of the attributes of the $ i^{\mathrm{th}} $ measurement. The initial idea of measuring the dissimilarity between $ \mathbf{O}^i $ and $ \mathbf{O}^j $ is by splitting them into three parts (\figPref\ref{fig:scheme_three_parts}), namely computing and weighting the dissimilarity of the shared and unshared attributes differently:
	\begin{equation}
		\label{eq:initNVDM}
		\begin{split}
			d_{CDM}(\mathbf{O}^i , \mathbf{O}^j )& = \sum_{a \in (A^i \cap A^j)}g({v}^i_a, {v}^j_a) +\\
			&\alpha \left( \sum_{a \in (A^i \backslash A^j)}g({v}^i_a, \gamma) + \sum_{a \in (A^j \backslash A^i)}g({v}^j_a, \gamma)\right), 
		\end{split}
	\end{equation}
	where $ g(\cdot, \cdot) $ is a chosen distance metric, $ {v}^i_a $ and $ {v}^j_a $ are the measured values associated with attribute $ a $. $ \gamma $ is a predefined value indicating a missing measured attribute and the regularization values $ \alpha \in \left[0, +\infty\right) $ are introduced to regulate or balance the contribution to the dissimilarity from those mutually unshared attributes. Herein we regulate the contribution of unshared attributes equally because there is no prior assumption that can be used to determine which of them should have more influence on the dissimilarity. In a specific application (e.g., \acl{fbp}), it might be reasonable to weight these two terms differently. The \acs{cdm} offers additional degrees of freedom owing to the contribution of the hyperparameters (\ie $ \gamma $ and $ \alpha $). Their values can be determined in a data driven approach according to the specific application.
	
	The basic \acs{cdm} formulated in \eqref{eq:initNVDM} weights the contribution from the shared and unshared attributes differently. However, we additionally propose two further variants of this measure which take also the actual numbers of these attributes into account. The application examples will later indicate that these are useful extensions. One is obtained by dividing the \acs{cdm} in \eqref{eq:initNVDM} by the total number of attributes thus yielding the average dissimilarity of the attributes, where $ |\cdot| $ denotes the cardinality of a set, here (The same symbol is used in this contribution also to indicate the absolute value of a scalar.): 
	
	\begin{equation}
		\label{eq:ADCM}
		d_{ACDM}(\mathbf{O}^i , \mathbf{O}^j ) = d_{CDM}(\mathbf{O}^i , \mathbf{O}^j )/|A_i\cup A_j|.
	\end{equation}
	We call this measure an \acf{acdm}. The second extension is obtained by weighting the terms in \eqref{eq:initNVDM} relatively according to the number of shared and unshared attributes, \ie
	\begin{equation}
		\label{eq:RDCM}
		\begin{split}
			&d_{RCDM}(\mathbf{O}^i , \mathbf{O}^j ) =  \sum_{a \in (A^i \cap A^j)}g({v}^i_a, {v}^j_a) + \\
			& \alpha\left(\omega_{A_i\backslash A_j} \sum_{a \in (A^i \backslash A^j)}g({v}^i_a, \gamma) + \omega_{A_j\backslash A_i}\sum_{a \in (A^j \backslash A^i)}g({v}^j_a, \gamma)\right)
		\end{split}
	\end{equation}
	where $ \omega_{A_i\backslash A_j} $ and $ \omega_{A_j\backslash A_i} $ are calculated by
	\begin{equation}
		\label{eq:rcdmWeight}
		\begin{split}
			& \omega_{A_i\backslash A_j} = \frac{|A_i\backslash A_j|}{(|A_i\cap A_j| + \epsilon)}\\
			& \omega_{A_j\backslash A_i} = \frac{|A_j\backslash A_i|}{(|A_i\cap A_j| + \epsilon)}
		\end{split}
	\end{equation}
	In \eqref{eq:rcdmWeight}, $ \mathcal{R}\ni \epsilon > 0 $ but $ \epsilon \ll 1 $ is introduced for avoiding division by zero in case there are no shared attributes at all. The \acf{rcdm} introduced in \eqref{eq:RDCM} yields a large dissimilarity value in such a case. Comparing to the widely used vector-based distance metrics, the \acsp{cdm} have three advantages:
	\begin{itemize}
		\item The measurements do not have to be rearranged into vectors of equal length.
		\item \acsp{cdm} can be used to balance the contributions to the dissimilarity from the shared and mutually unshared attributes.
		\item \acsp{cdm} have hyperparameters and are capable of adapting to different data.
	\end{itemize}
	
	Subsequently, we compare the three proposed \acsp{cdm} by applying them to \acs{fbp} using different values of the hyperparameters and joining them with selected distance metrics (see \secPref \ref{sec:experimentalResults}).
	%AW: the story about the 40/50 metrics and chosing the 8 best ones has alredy been told above - avoid repetition
	
	\renewcommand{\arraystretch}{1.75}
	\begin{table}[!htb]
		\centering
		\caption{Distance metrics used herein}
		\label{tab:metrics}
		\begin{tabular}{cc}
			\hline
			Metric & Equation\footnote{The formula and nomenclature are from \cite{cha2007comprehensive} except the formula of Jaccard distance. In all these equations $ \mathbf{x}_i $ and $ \mathbf{y}_i $ are the $ i $-th element of the vectors $ \mathbf{x}$ and $ \mathbf{y} $, respectively and $ d $ is the dimension of $ \mathbf{x}$ and $ \mathbf{y} $.}\\
			\hline
			Lorentzian & $ g_{Lor}( \mathbf{x},  \mathbf{y}) = \sum_{i=1}^{d}\ln (1 + |\mathbf{x}_i - \mathbf{y}_i|)$ \\ 
			Hamming \footnote{$ I(\cdot) $ is an indicator function and it yields 1 if and only if the condition is fulfilled.} & $ g_{Ham}(\mathbf{x},  \mathbf{y})=\sum_{i=1}^{d}I(\mathbf{x}_i \neq \mathbf{y}_i)/d $\\
			Jaccard \footnote{This formula is taken from \cite{8115922}. $ \gamma $ is the indicator of a missing measured attribute. In case of \acs{cdm}, Hamming and Jaccard distances are equivalent.}&$  g_{Jac}(\mathbf{x},  \mathbf{y})= \frac{\sum_{i=1}^{d}I((\mathbf{x}_i  \neq \mathbf{y}_i )\, \& \,\left[(\mathbf{x}_i \neq \gamma)\, | \,(\mathbf{y}_i \neq \gamma)\right])}{\sum_{i=1}^{d}I((\mathbf{x}_i \neq \gamma) \,|\, (\mathbf{y}_i \neq \gamma))}$\\
			Wave Hedges&$ g_{WH}(\mathbf{x},  \mathbf{y})= \sum_{i=1}^{d}(|\mathbf{x}_i - \mathbf{y}_i|/\max(|\mathbf{x}_i|, |\mathbf{y}_i|))$\\
			Canberra&$ g_{Can}(\mathbf{x},  \mathbf{y})= \sum_{i=1}^{d}(|\mathbf{x}_i - \mathbf{y}_i|/(|\mathbf{x}_i| + |\mathbf{y}_i|))$\\
			Clark&$  g_{Cla}(\mathbf{x},  \mathbf{y})=\sqrt{\sum_{i=1}^{d}(|\mathbf{x}_i - \mathbf{y}_i|/(|\mathbf{x}_i| + |\mathbf{y}_i|))^2}$\\
			City block \footnote{In \cite{6418937}, City Block is named Manhattan distance.}&$ g_{CB}(\mathbf{x},  \mathbf{y})= \sum_{i=1}^{d} |\mathbf{x}_i - \mathbf{y}_i|$\\
			Minkowski&$ g_{Min}(\mathbf{x},  \mathbf{y}, p)= \sqrt[p]{\sum_{i=1}^{d} |\mathbf{x}_i - \mathbf{y}_i|^p}$\\
			\hline
		\end{tabular}
	\end{table}
	
	\section{An application of \acsp{cdm} to \acs{fbp}}\label{sec:experimentalResults}
	In this section, we first describe the fundamentals of \acs{fbp}, the widely used positioning algorithm \acs{knn}, the chosen evaluation metrics, and four datasets used for practical application and assessment. We then evaluate the performance of the three \acs{cdm} in terms of positioning results taking into account only very few different values $ \alpha $ . Then  the \acf{cv} method is applied to search for particularly suitable values of $ \alpha $ for a chosen distance metric and dataset. We compare the positioning performance of the approach using the \acs{cdm} to that of \acs{knn} without \acs{cdm}.
	
	\subsection{The baseline algorithm, performance criteria and data sets}\label{subsec:testbed}
	% AW: changed order as indicated before - the data are just used for demonstration and should not be presented before the aglgorithms
	
	\subsubsection{Fingerprinting-based positioning}
	The measured fingerprints for the purpose of \acs{fbp} have the characteristic of missing attributes because the coverage of an \acs{ap} is restricted by the transmitting  power, free space loss, signal attenuation and the sensitivity of the receiver. The coverage is higher for higher transmission power, higher sensitivity and lower attenuation. One benefit of using \acs{cdm} instead of vector-based distance metrics is that it avoids the need for conversion of the measurements into vectors of equal length. Further experimental analysis in the consecutive sections shows that using \acs{cdm} can also improve the positioning accuracy and stability.
	
	Generally an \acs{fbp} (\eg using the signal strength from \acs{wlan} \acsp{ap} as the fingerprints for indoor positioning) consists of two stages: offline fingerprinting stage and online positioning stage. During the offline stage, a \acs{rfm} $ \{\left({\mathbf{O}^i, \mathbf{l}^i}\right)\}_{i=1}^{N} $ representing the relationship between the measurements (\eg \acs{rss}) and locations  is collected via carrying out a site survey (either by a professional surveyor or by crowd-sourcing). During online positioning stage a user measured observation $ \mathbf{O}^{\mathrm{u}} $ is matched to the \acs{rfm} using a \acs{fbp} algorithm $ f $ for estimating the user's location $ {\mathbf{l}}^{\mathrm{u}}$ (an estimated location is denoted as $ \hat{\mathbf{l}}^{\mathrm{u}}$), \ie $ f:  \mathbf{O}^{\mathrm{u}}\rightarrow \hat{\mathbf{l}}^{\mathrm{u}}$.
	
	Herein we use \acs{knn}, one of the most widely used \acs{fbp} algorithms, as the baseline positioning method for evaluation and comparison. \textit{UJIIndoorLoc} (see \secPref\ref{sec:Datasets}) is a dataset including multiple buildings and multiple floors. We use the hierarchical \acs{knn} according to \cite{7275492} for processing this dataset. For the other datasets we use \acs{knn} as follows:
	\begin{itemize}
		\item Computing the dissimilarity measure between the user's measurement and the ones stored in the \acs{rfm};
		\item Finding the $ k $ nearest reference points in the feature space, \ie reference points with the $ k $ smallest dissimilarity;
		\item Taking the average coordinates of these $ k $ reference points as the user's location.
	\end{itemize}
	More details about \acs{fbp} and \acs{knn} can be found, \eg in \cite{ Zhou2018, Padmanabhan2000}.
	\subsubsection{Evaluation of positioning performance}
	The Euclidean distance between the estimated location $ \hat{\mathbf{l}} $ and the ground truth location  $ \mathbf{l} $ is used as the basic evaluation of the positioning error. In addition, we also use the statistical values (\eg mean or standard deviation), \acf{rmse}, and \acf{ecdf} with respect to the error distance as further performance evaluation.
	
	The implementation of the proposed \acsp{cdm} and their relevant functions are in Python and partially based on the scikit-learn package \cite{scikit-learn}.
	
	\subsubsection{Available positioning datasets}\label{sec:Datasets}
	We use four different datasets (both the \acsp{rfm} and validation datasets) from fingerprinting-based \acs{wlan} indoor positioning systems for evaluating and comparing the performance of the proposed \acsp{cdm}. The datasets \textit{Alcala2017}, \textit{Tampere} and \textit{UJIIndoorLoc} are available online \cite{Sansano2016} and more details about them can be found in \cite{TORRESSOSPEDRA20159263,7275492}. Another dataset \textit{HIL} is described in \cite{Zhou2018}. These four datasets represent different \acs{fbp} scenarios with respect to area of indoor region, number of available \acsp{ap}, method of fingerprint collection, and device heterogeneity. The summarized characteristics of the datasets are given in \tabPref \ref{tab:summaryRFMs}.

	%AW: Please align the numbers in this table on the right hand side, not the left such that digits with equal value are printed below each other
	
	\renewcommand{\arraystretch}{1.15}
	\begin{table}[!htb]
		\centering
		\caption{The characteristics of the datasets (Tampere: herein only data from one building; UJIIndoorLoc: clearing procedure applied see \nameref{sec:appendix})}
		\small
		\label{tab:summaryRFMs}
%		\begin{tabular}{p{0.22\columnwidth}p{0.07\columnwidth}rp{0.06\columnwidth}rr}
		\begin{tabular}{>{\centering}p{0.22\columnwidth}>{\centering}p{0.07\columnwidth}>{\raggedleft\arraybackslash}p{0.09\columnwidth}>{\centering}p{0.06\columnwidth}>{\centering}p{0.13\columnwidth}>{\raggedleft\arraybackslash}p{0.13\columnwidth}}
			\noalign{\smallskip}\hline\noalign{\smallskip}
			\hspace{1.5ex}
			\multirow{2}{*}{Dataset} & \multirow{2}{*}{\shortstack{$ \mathrm{\sharp} $ Buil-\\dings}} &\multirow{2}{*}{$ \mathrm{\sharp} $ Floors} &\multirow{2}{*}{$ \mathrm{\sharp} $\acsp{ap}}  & \multirow{2}{*}{\shortstack{$ \mathrm{\sharp} $ Training\\ samples}} & \multirow{2}{*}{\shortstack{$ \mathrm{\sharp} $ Validation \\samples\footnote{In case there is no provided validation samples, we randomly split the training samples into two datasets. 75\% of them are used for training and the remaining ones are used for validation.}}} \tabularnewline\tabularnewline
%			\hline
			\noalign{\smallskip}\hline\noalign{\smallskip}
			\textit{Alcala2017}&1&1&152&670&0\tabularnewline
			\textit{HIL}&1&1&490&1525&509\tabularnewline
			\textit{Tampere} &1&4&309&1478&0\tabularnewline
			\textit{UJIIndoorLoc}&3&4--5&520&3818&1110\tabularnewline
			\noalign{\smallskip}\hline\noalign{\smallskip}
		\end{tabular}
	\end{table}
	In case of applying the proposed \acsp{cdm} to \acs{fbp}, we use a fixed value of $ \gamma $ for indicating the missing attributes. In \textit{HIL}, $ \gamma$ is set to -110 dBm and $ \gamma=100 $ in other three datasets \cite{7275492}.
	
	\subsection{Evaluation  and comparison of different \acsp{cdm}}
	Herein we propose three versions to \acsp{cdm}. However, we want to briefly investigate whether \acs{fbp} is less sensitive with respect to dataset and distance metric included than the others. Since the results may also depend on $ \alpha $, we use a few fixed values  \ie $ \alpha = \{0.5, 1.0, 1.25\} $ for the analysis and compare the results.
	
	\begin{figure}[!htb]
		\centering
		\subfloat[\textit{Alcala2017}]{
			\label{subfig:three_cdms_acala}
			\includegraphics[width=0.45\columnwidth]{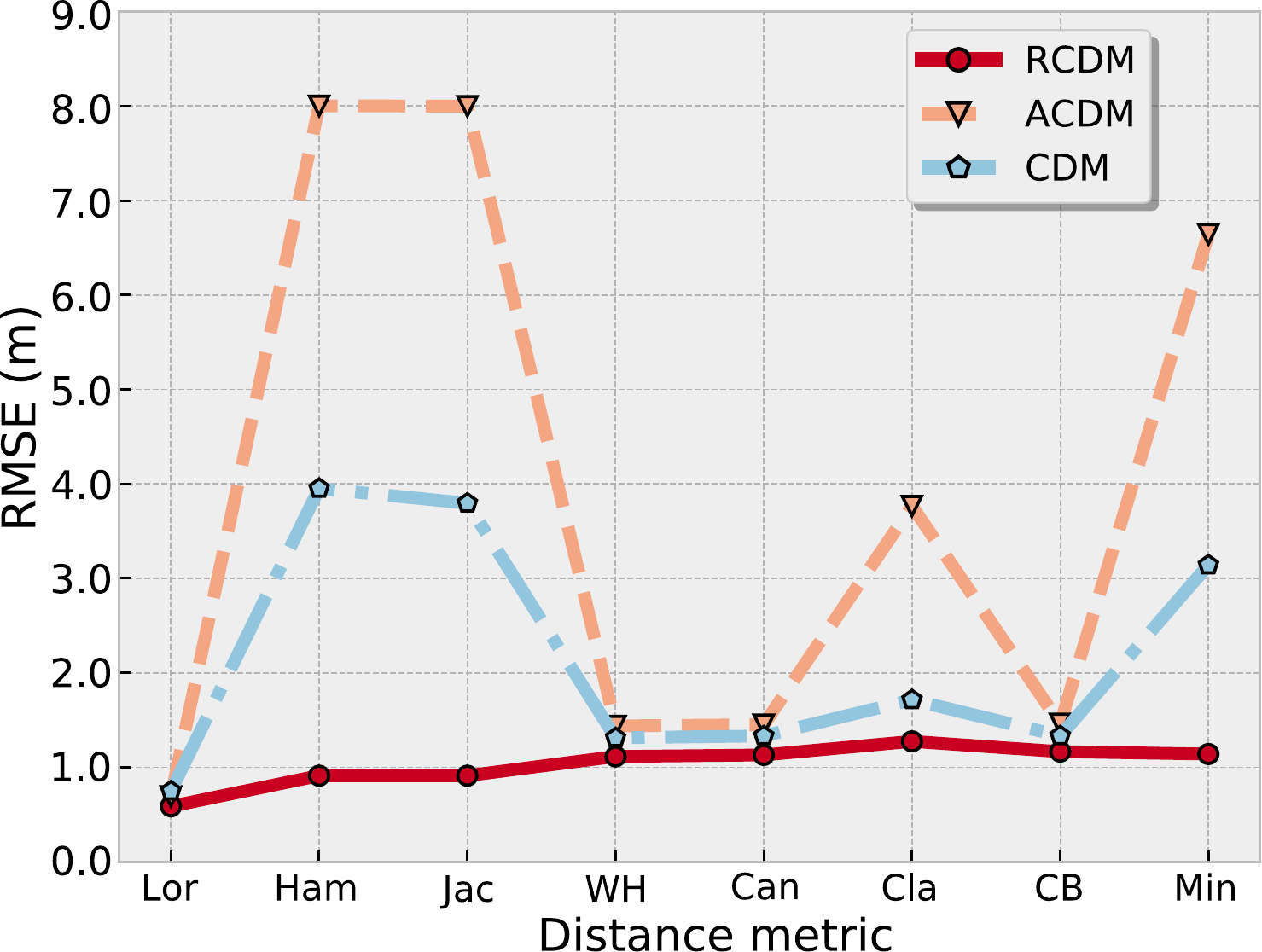}}\hspace{1ex}
		\subfloat[\textit{HIL}]{
			\label{subfig:three_cdms_hil}
			\includegraphics[width=0.45\columnwidth]{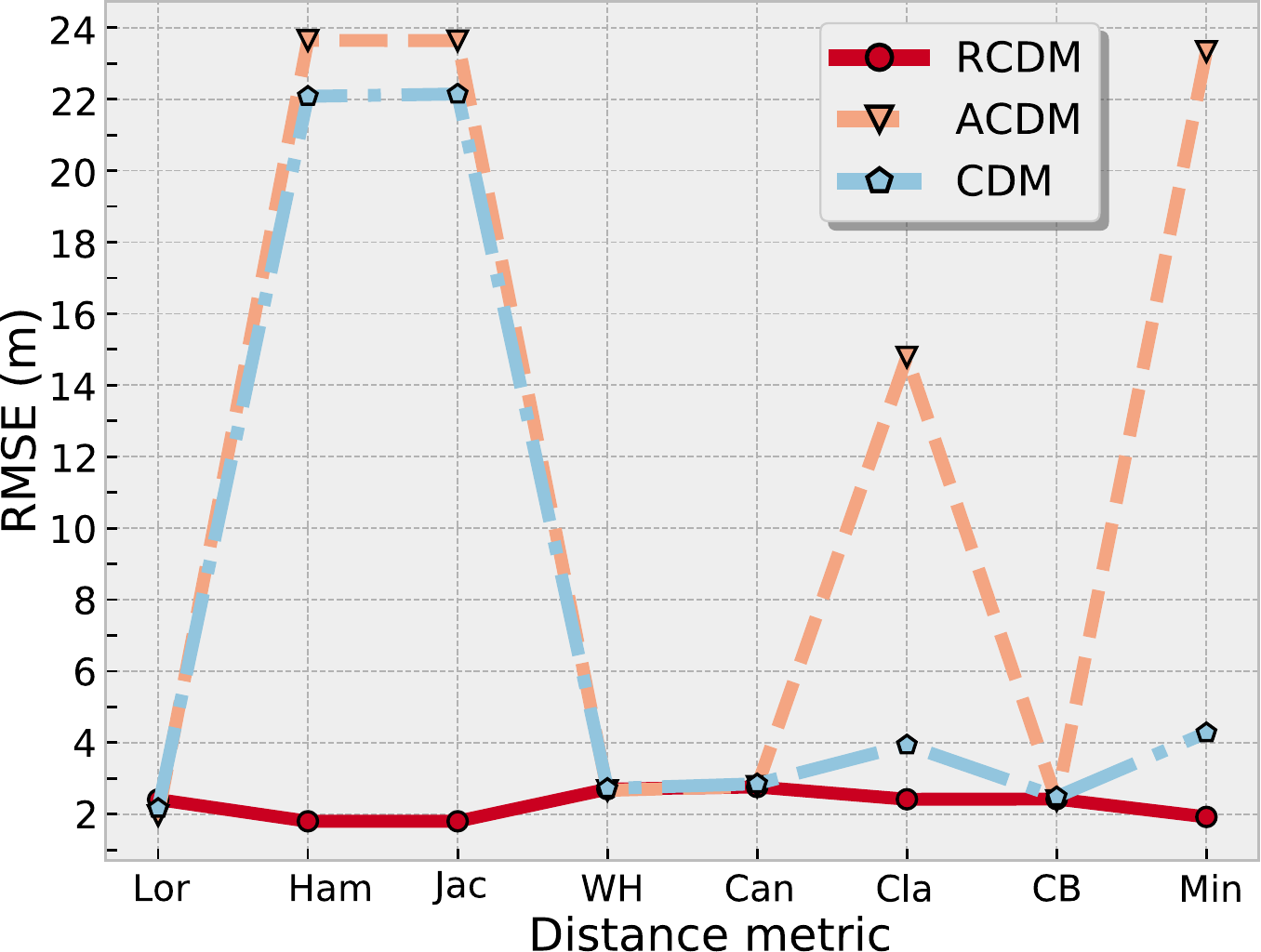}}\\
		\subfloat[\textit{Tampere}]{
			\label{subfig:three_cdms_tampere}
			\includegraphics[width=0.45\columnwidth]{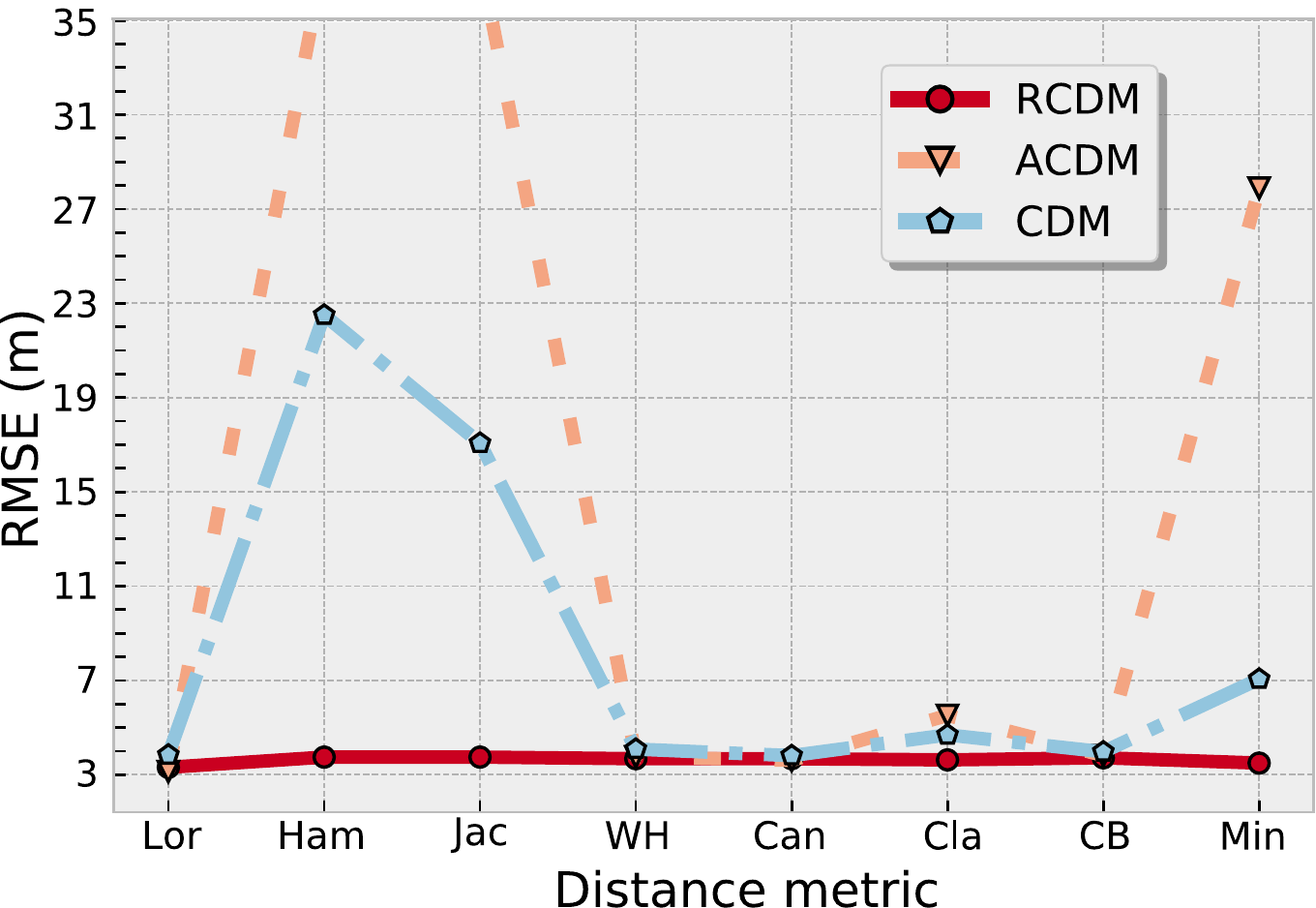}}\hspace{1ex}
		\subfloat[\textit{UJIIndoorLoc}]{
			\label{subfig:three_cdms_uji}
			\includegraphics[width=0.45\columnwidth]{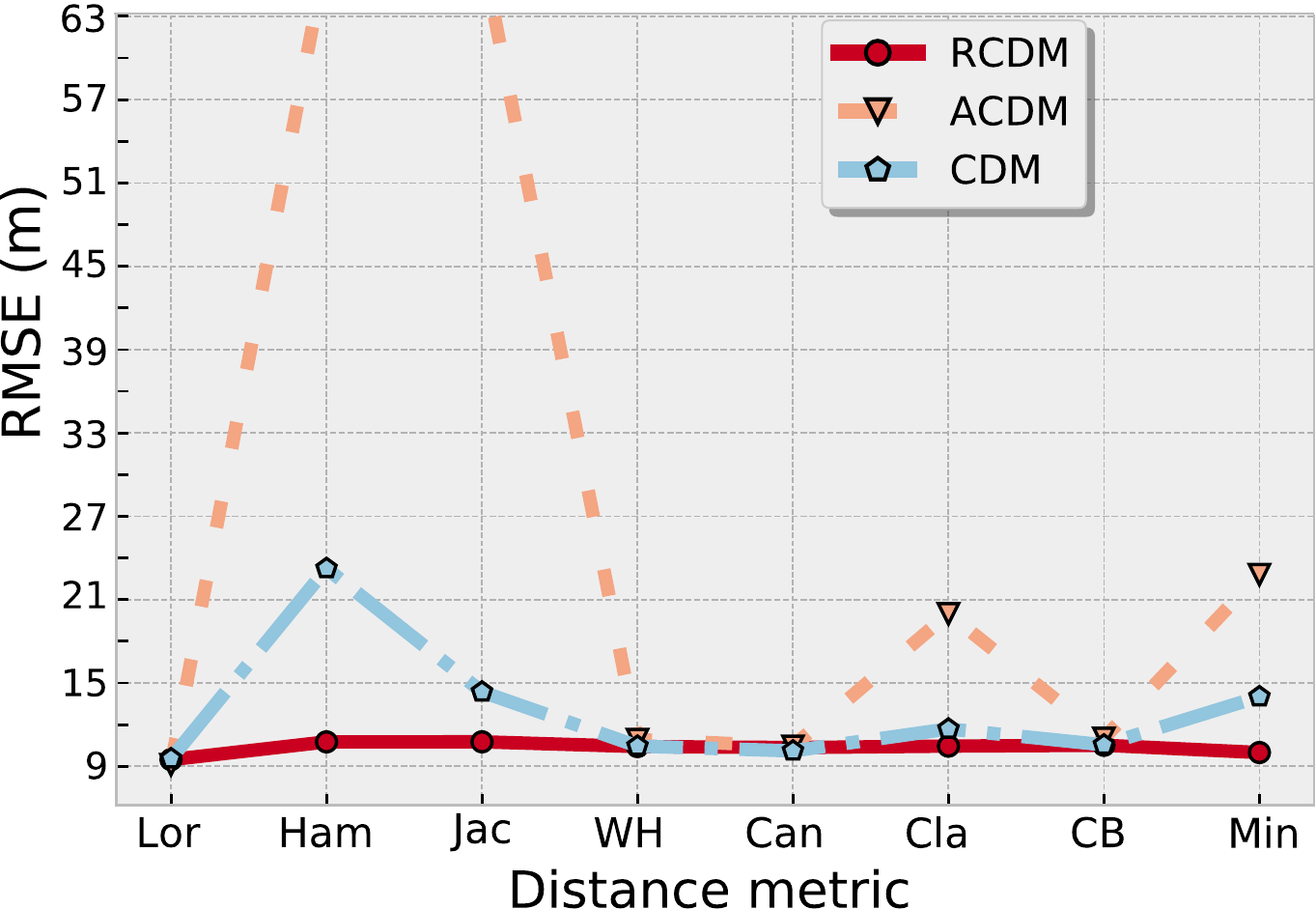}}
		\caption{An example of \acs{rmse} of three \acsp{cdm} with $ \alpha = 1.0 $}
		\label{fig:rmse_three_cdms}
	\end{figure}
	\figPref \ref{fig:rmse_three_cdms} shows the \acs{rmse} of the positioning result using all three \acsp{cdm} with $ \alpha = 1.0 $. Similar results are also obtained using the other two values of $ \alpha $. From \figPref \ref{fig:rmse_three_cdms} , we can conclude that the \acs{cdm} relatively weighted by the shared and mutually unshared attributes performs more stable than that of other two \acsp{cdm} in the sense that the \acs{rmse} of all four datasets of using \acs{rcdm} compounding with all eight distance metrics  has the smallest deviation. Therefore, we focus on the application of \acs{rcdm} for the remainder of this paper.
	
	\subsection{Tuning the regularization value $ \alpha $}
	In order to find suitable values of $ \alpha $ we carry out \acf{cv} \cite{Bishop:2006:PRM:1162264}, a widely used method for model selection, for various distance metrics and the four datasets. \acs{cv} can make full use of the training dataset by randomly splitting it into several folds and iteratively using one of them as the temporal test samples and the remaining ones as the temporal training dataset. Herein we use 10-fold \acs{cv} and search for the suitable value of $\alpha$ in the range of $ [0, 3] $ with the interval of 0.1. In this paper, we only illustrate that a useful value of $ \alpha $ can be found using \acs{cv} without specifically looking for the most appropriate search space of $ \alpha $. The related investigation into search space and optimal value is left for future work. In the consecutive part of this section, we mainly show the results of \textit{HIL} and \textit{UJIIndoorLoc} using \acs{knn} ($ k=1 $) as the \acs{fbp} algorithm with \acs{rcdm} (compounding with Lorentzian and Minkowski) for remaining the clarity. The results of other datasets and distance metrics are also similar to what presented herein.
	
	\begin{figure}[!htb]
		\centering
		\subfloat[Lorentzian]{
			\label{subfig:cv_hil_min_k_3}
			\includegraphics[width=0.618\columnwidth]{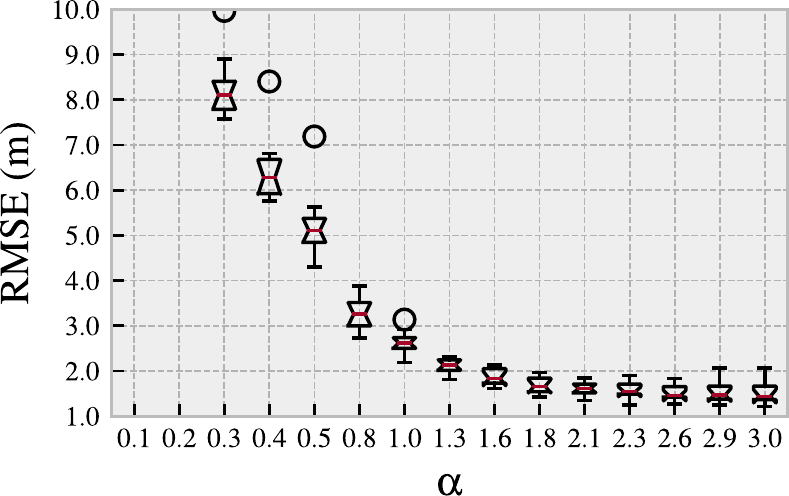}}\\%\hspace{1ex}
		\subfloat[Minkowski]{
			\label{subfig:cv_tamp_can_k_1}
			\includegraphics[width=0.618\columnwidth]{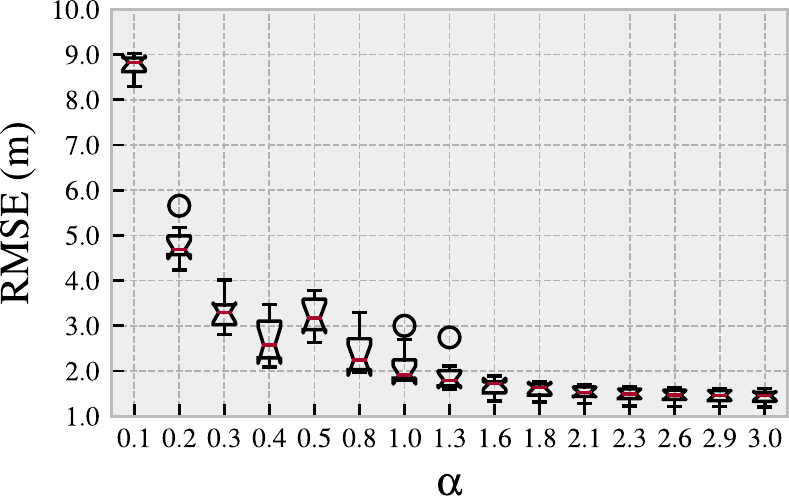}}
		\caption{\acs{cv} box-plots of \textit{HIL} ($ k=1 $ , \acs{rcdm}). }
		\label{fig:cv_alpha}
	\end{figure}
	\begin{figure}[!htb]
		\centering
		\subfloat[Lorentzian]{
			\label{subfig:cv_uji_lor_k_1}
			\includegraphics[width=0.618\columnwidth]{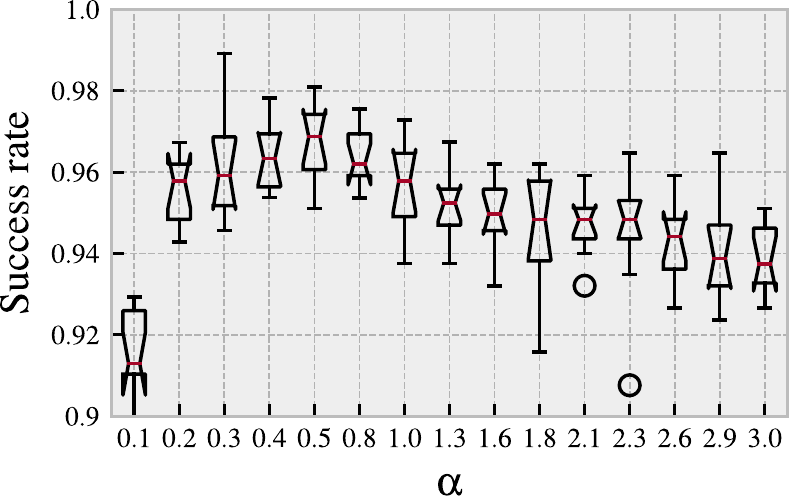}}\\%\hspace{1ex}
		\subfloat[Minkowski]{
			\label{subfig:cv_uji_min_k_1}
			\includegraphics[width=0.618\columnwidth]{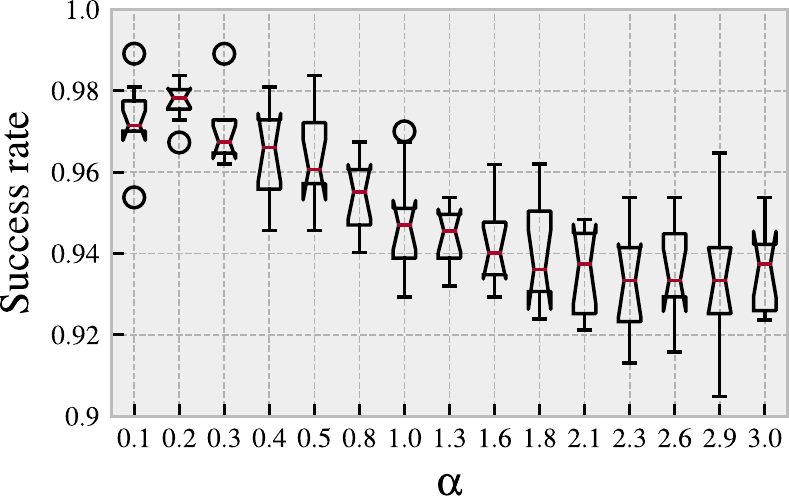}}
		\caption{\acs{cv} box-plots of \textit{UJIIndoorLoc} ($ k=1 $ , \acs{rcdm}). }
		\label{fig:cv_alpha_uji}
	\end{figure}
	
	As shown in \figPref \ref{fig:cv_alpha} and \figPref\ref{fig:cv_alpha_uji}, there is a value of $ \alpha $ resulting in the minimum \acs{rmse} and maximum success rate for a chosen distance metric and dataset. Regarding dataset \textit{Alcala2017}, \textit{HIL}, and \textit{Tampere}, we select the $ \alpha $ which achieves minimum average value of \acs{rmse} of 10-fold \acs{cv} as the suitable value for a chosen distance metric and dataset. We only plot part of the cross validation result for preserving the clarity. For \textit{UJIIndoorLoc}, we use the regularization value which obtains the maximum average success rate (\figPref\ref{fig:cv_alpha_uji}), defined as the percentage of correctly locating both the building and floor, as the proper value of $ \alpha $, since in case of applying \acs{fbp} to multi-buildings and multi-floors, the success rate is a better indicator for the positioning performance than using \acs{rmse} \cite{6418937}. One reason using success rate instead of \acs{rmse} as the criterion is that the wrongly locating either the buildings or the floors introduces large positioning errors and it makes that the \acs{rmse} is no longer a good indicator of positioning performance. From the \acs{cv} results shown in  \figPref \ref{fig:cv_alpha} and \figPref\ref{fig:cv_alpha_uji}, the suitable values of $ \alpha $ for \textit{HIL} and \textit{UJIIndoorLoc} are 2.7 and 3.0, and 0.5 and 0.2 in case of relatively compounding with both Lorentzian and Minkowski ($ p=2 $) distances, respectively.
	
	\subsection{Comparison of positioning performance}
\begin{figure}[!htb]
		\centering
		\subfloat[\textit{Alcala}]{
			\label{subfig:cmp_alcala_k_1}
			\includegraphics[width=0.618\columnwidth]{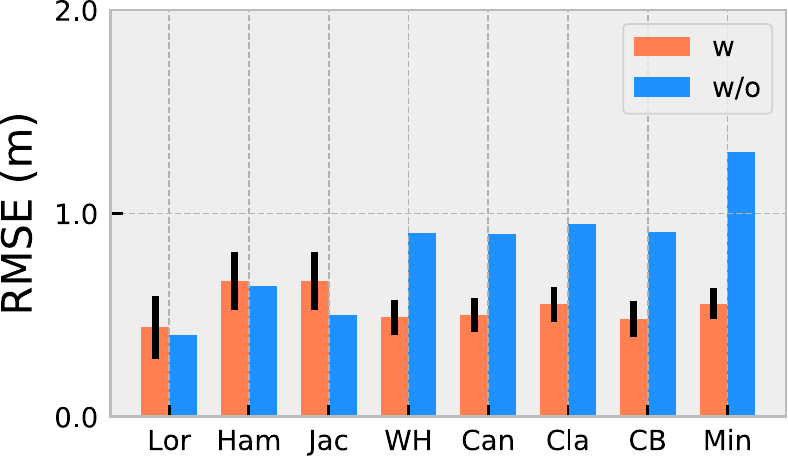}}\\
		\subfloat[\textit{HIL}]{
			\label{subfig:cmp_hil_k_1}
			\includegraphics[width=0.618\columnwidth]{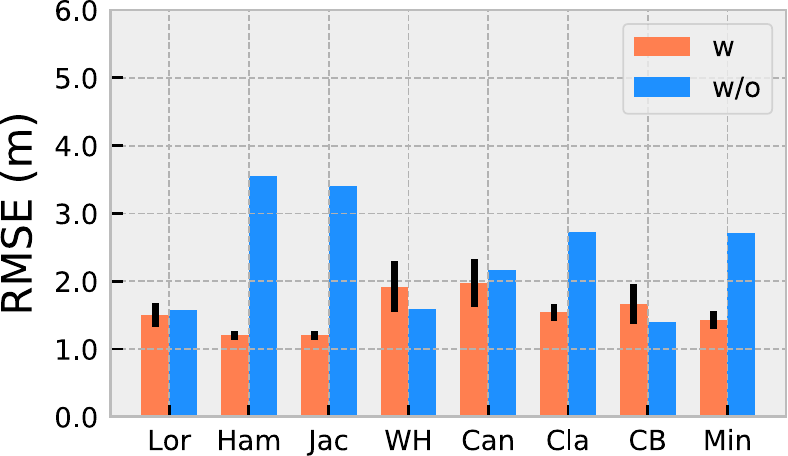}}\\
		\subfloat[\textit{Tampere}]{
			\label{subfig:cmp_tampere_k_1}
			\includegraphics[width=0.618\columnwidth]{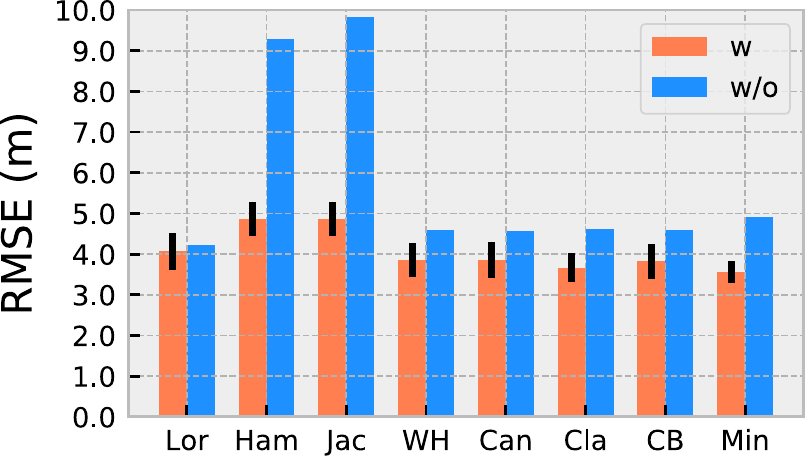}}
		\caption{Comparison of \acs{rmse} ($ k=1 $, \acs{rcdm}). In the figure, w and w/o denotes the ones with \acs{rcdm} and  without \acs{rcdm}, respectively.}
		\label{fig:cmp_rmse_datasets}
	\end{figure}

\begin{figure}
		\centering
		\subfloat[]{
			\label{subfig:cmp_facc_uji_k_1}
			\includegraphics[width=0.618\columnwidth]{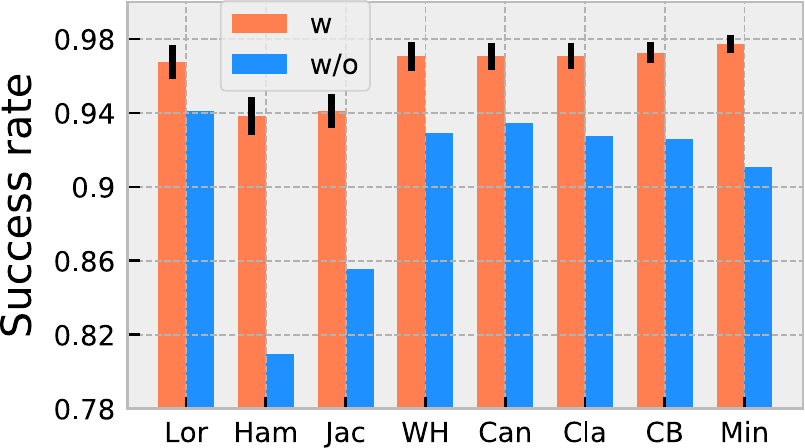}}\\%\hspace{1ex}
		\subfloat[]{
			\label{subfig:cmp_rmse_uji_k_1}
			\includegraphics[width=0.618\columnwidth]{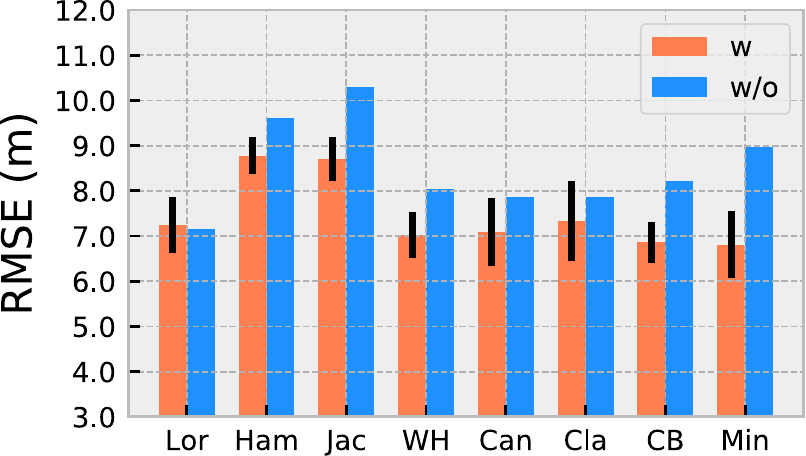}}
		\caption{Comparison of success rate and \acs{rmse} of  \textit{UJIIndoorLoc} ($ k=1 $). }
		\label{fig:cmp_rmse_facc_datasets_uji}
	\end{figure}

	We compare the \acs{rmse} of the positioning result obtained using \acs{rcdm} (using the regularization value ($ \alpha $) found by \acs{cv}) to the ones attained using vector-based distance metrics (\figPref\ref{fig:cmp_rmse_datasets}). As shown in \figPref\ref{fig:cmp_rmse_datasets} and \figPref\ref{subfig:cmp_rmse_uji_k_1}, the proposed \acs{rcdm} outperforms almost all eight original distance metrics on all four datasets (except in case of compounding with  Hamming and Jaccard distances on \textit{Alcala2017} (see \figPref\ref{subfig:cmp_alcala_k_1}) and Wave Hedges and city block distances on \textit{HIL} (see \figPref\ref{subfig:cmp_hil_k_1}). In addition, the reduction of the \acs{rmse} is over two times comparing to that of without using \acs{rcdm} and the deviation of the \acs{rmse} of compounding with all eight distance metrics is much smaller than that of the original ones. In \figPref\ref{fig:cmp_rmse_facc_datasets_uji}, the success rate of using \acs{rcdm} is higher than that of using the original distance metrics and the improvement is up to 13\% (\figPref\ref{subfig:cmp_facc_uji_k_1}). 
	\begin{table*}
		\centering
		\caption{Positioning results of \textit{UJIIndoorLoc}}
		\label{tab:results_uji}
		\begin{tabular}{cccccccccc}
			\noalign{\smallskip}\hline\noalign{\smallskip}
			Metrics & & Lor&Ham & Jac & WH & Can & Cla & CB & Min\\
			\noalign{\smallskip}\hline\noalign{\smallskip}
			\multirow{2}{*}{\shortstack{Building \\accuracy (\%)}}&w/o&99.76&96.65&99.67&99.92&99.92&99.92&99.92&99.84\\
			& w&99.92&99.92&99.92&99.92&99.92&99.92&99.92&99.92\\
			\noalign{\smallskip}\hline\noalign{\smallskip}
			\multirow{2}{*}{\shortstack{Success \\rate (\%)}}&w/o&94.12&80.98&85.55&92.9&93.47&92.73&92.57&91.1\\
			& w&96.33&93.71&93.71&97.22&96.73&96.24&97.47&96.98\\
			\noalign{\smallskip}\hline\noalign{\smallskip}
			\multirow{2}{*}{\shortstack{Median of\\error distance ($ \text{m} $)}}&w/o&1.78&2.04&1.73&2.76&2.48&2.65&2.89&3.92\\
			& w&1.44&2.07&2.07&1.48&1.53&1.64&1.38&1.47\\
			\noalign{\smallskip}\hline\noalign{\smallskip}
			\multirow{2}{*}{\shortstack{80-percentile of \\error distance ($ \text{m} $)}}&w/o&9.35&14.72&14.76&11.06&10.65&10.65&11.19&13.76\\
			& w&8.36&11.58&11.58&8.12&8.34&8.79&7.96&8.11\\
			\noalign{\smallskip}\hline\noalign{\smallskip}
		\end{tabular}
	\end{table*}
	
\begin{figure}
		\centering
		\subfloat[Without \acs{rcdm}]{
			\label{subfig:cmp_ecdf_w_o_cdm_k_1}
			\includegraphics[width=0.618\columnwidth]{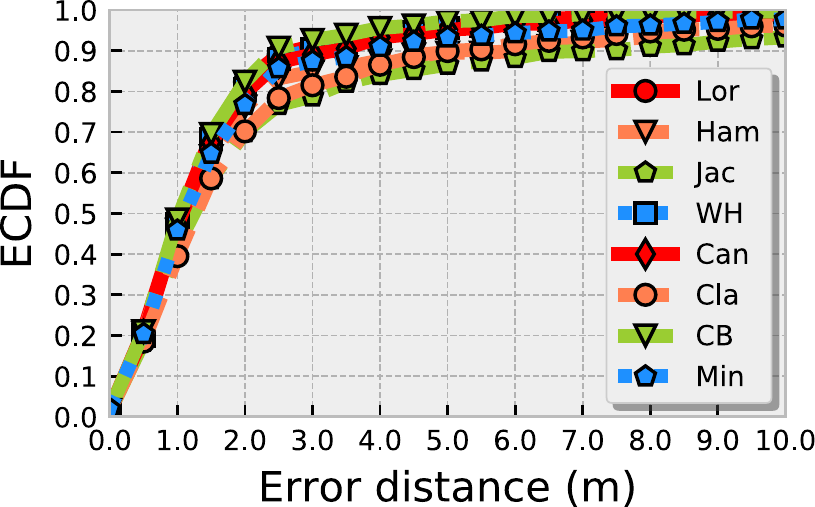}}\\%\hspace{0.2ex}
		\subfloat[With \acs{rcdm}]{
			\label{subfig:cmp_ecdf_w_cdm_k_1}
			\includegraphics[width=0.618\columnwidth]{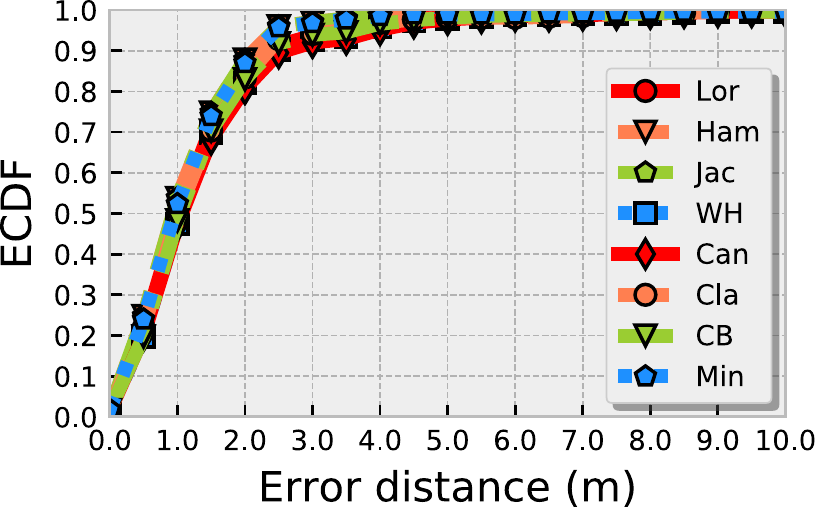}}
		\caption{Comparison of \acs{ecdf} between different distance metrics (\textit{HIL}, $ k=1 $)}
		\label{fig:cmp_ecdf_hil_all_8}
	\end{figure}

	According to the comparison of the \acf{ecdf} of all eight original distance metrics and \acs{rcdm} of dataset \textit{HIL} (\figPref\ref{fig:cmp_ecdf_hil_all_8}), we can conclude that the cumulative positioning accuracy using \acs{rcdm} is higher than that of using original distance metrics. In addition, the maximum positioning error distance using \acs{rcdm} is much smaller than that without using \acs{rcdm}. The \acs{fbp} algorithms achieving a small maximum positioning error distance make it easier to constrain the upper bound of the positioning error.
	
	In \tabPref\ref{tab:results_uji}, we present the positioning results of applying the proposed approach to \textit{UJIIndoorLoc}. From the building accuracy, defined as the percentage of correctly identifying the building, it seems that there are some validation samples which are not placed in the correct building by the positioning algorithm, because the building accuracy of \acs{rcdm} compounding with different distance metrics keeps the same, \ie the building accuracy is saturated. This saturation might be caused by the cleaning of the dataset (see \nameref{sec:appendix}). Regarding the success rate, it improves about 5\% on average using the \acs{rcdm} and is over 97\% of correctly identifying both the building and floor. In addition, the median and 80-percentile positioning error distances using \acs{rcdm} reduce obviously except compounding with Hamming and Jaccard distances.
	
	\section{Conclusion}\label{sec:futureWork}
	We propose a non-vector-based dissimilarity measure, named \acf{cdm}, by combining a typical distance metric with set operations for the purpose of measuring the dissimilarity between measurements despite the possibility of missing attributes. The proposed \acs{cdm} is flexible because it includes hyperparameters, which can be tuned according to the data and needs of the application. We apply the proposed \acs{cdm} to four datasets collected in fingerprinting-based \acs{wlan} indoor positioning systems and the positioning performance verifies the validity of it. Both the accuracy of identifying buildings and floors, and the specific locations improve obviously, which are over 5\% and 10\%, respectively. Although the \acs{cdm} is proposed herein starting from the idea of handling missing data in \acl{fbp}, it is applicable to other missing data problems as well (\eg searching the correspondences of point clouds according to sparsely described local features).
	
	\section*{Acknowledgment}
	{The China Scholarship Council  (CSC) financially supports the first author's doctoral research.}
	\section*{Appendix}\label{sec:appendix}
	We clean the \textit{UJIIndoorLoc} dataset from two aspects: i) the invalid samples, and ii) the replicas in the training dataset. An invalid sample is a measurement that all of the  \acsp{ap} are filled with missing values. A replica of the measurement is that at least two measurements are measured at the same location by the same user using the same device in a short time range (\eg less than 5 minutes).
	\begin{itemize}
		\item Invalid samples: We find that 76 out of 20013 samples in the training dataset are invalid measurements by checking whether the \acs{rss} of all \acsp{ap} of a measurement is indicated by a missing value (\eg 100 used in \textit{UJIIndoorLoc}). We thus delete them from the training dataset.
		\item Replicas: These replicas are highly correlated and they might cause the failure of the cross validation using the training dataset because it is easy to get an over-optimistic results using a dataset containing replicas for cross validation \cite{Bishop:2006:PRM:1162264}. This makes the parameter found by cross validation not applicable to another test dataset. We find out there are a lot of replicas in the training dataset (only 3818 out of 19937 reference measurements do not have a replica). We thus randomly sample  one of those replicas as the reference fingerprint for the training dataset in our experimental analysis \footnote{Herein we use only one of the replicas as the reference fingerprint, however, it is useful that grouping or averaging those replicas as one reference fingerprint.}.
	\end{itemize}
	\bibliographystyle{IEEEtran}
	\bibliography{../reference/ipin_2018}
	%\newpage

\end{document}

%% file: cz_acronyms.tex
%definition all acronymous
%\usepackage{acronym}
\acrodef{vc}[VC]{Vapnik-Chervonenkis}
\acrodef{knn}[$ k $NN]{$ k $ nearest neighbors}
\acrodef{lbs}[LBS]{location-based service}
\acrodef{ilbs}[ILBS]{indoor location-based service}
\acrodef{fips}[FIPS]{fingerprinting-based indoor positioning system}
\acrodef{fbp}[FbP]{fingerprinting-based positioning}
\acrodef{gnss}[GNSS]{global navigation satellite system}
\acrodef{ips}[IPS]{indoor positioning system}
\acrodef{rfid}[RFID]{radio frequency identification}
\acrodef{uwb}[UWB]{ultra wideband}
\acrodef{wlan}[WLAN]{wireless local area network}
\acrodef{rss}[RSS]{received signal strength}
\acrodef{ap}[AP]{access point}
\acrodef{roi}[RoI]{region of interest}
\acrodef{lasso}[LASSO]{least absolute shrinkage and selection operator}
\acrodef{rfm}[RFM]{reference fingerprint map}
\acrodef{map}[MAP]{maximum a posteriori}
\acrodef{cpa}[CPA]{cumulative positioning accuracy}
\acrodef{mse}[MSE]{mean squared error}
\acrodef{tp}[TP]{test position}
\acrodef{wrt}[w.r.t.]{with respect to}
\acrodef{mac}[MAC]{media access control}
\acrodef{imu}[IMU]{inertial measurement unit}
\acrodef{foba}[adaFoBa]{adaptive forward-backward greedy}
\acrodef{rp}[RP]{reference point}
\acrodef{mji}[MJI]{modified Jaccard index}
\acrodef{ble}[BLE]{Bluetooth low energy}
\acrodef{oil}[OIL]{organic indoor localization}
\acrodef{will}[WILL]{wireless indoor localization without site survey}
\acrodef{hiwl}[HIWL]{hidden Morkov model-based indoor wireless localization}
\acrodef{svm}[SVM]{support vector machine}
\acrodef{lda}[LDA]{linear discriminant analysis}
\acrodef{sop}[SoP]{signal of opportunity}
\acrodef{ks}[KS]{Kolmogorov-Smirnov}
\acrodef{ecdf}[ECDF]{empirical cumulative distribution function}
\acrodef{cdm}[CDM]{compound dissimilarity measure}
\acrodef{rmse}[RMSE]{root mean squared error}
\acrodef{rcdm}[RCDM]{relatively weighted compound dissimilarity measure}
\acrodef{acdm}[ACDM]{averagely weighted compound dissimilarity measure}
\acrodef{cv}[CV]{cross validation}
\acrodef{pdf}[PDF]{probability density function}

%% file: cz_command.tex
\newcommand{\todo}[1]{\hl{\textit{Tbd: #1}}}
\newcommand{\secPref}{Section }
\newcommand{\figPref}{Fig. }
\newcommand{\tabPref}{TABLE }
\newcommand{\eg}{e.g., }
\newcommand{\ie}{i.e. }
\newcommand{\etal}{et al. }
\newcommand{\rhl}[1]{\textcolor{red}{\hl{#1}}}